%% file: revision9.tex
\documentclass[journal,11pt,draftclsnofoot,onecolumn]{IEEEtran}

\usepackage{tikz}
\usetikzlibrary{matrix,arrows}

\usepackage{amsmath,amssymb,mathrsfs,amsthm}
\usepackage[colorlinks=true,linkcolor=black,anchorcolor=black,citecolor=black,filecolor=black,menucolor=black,urlcolor=black]{hyperref}
\usepackage{graphicx}
\usepackage{psfrag}
\usepackage{subfigure}
\usepackage{cite}
\usepackage{multirow}
\usepackage{stfloats}
\usepackage{epsfig}
\usepackage{amsfonts}
\usepackage{graphics}
\usepackage{color}
\usepackage{mathtools}
\DeclarePairedDelimiter{\ceil}{\lceil}{\rceil}

\epsfverbosetrue

\include{defns}  

\newcommand{\lexprod}{\circ}

\def\0{\bf{0}}

\newtheorem{theorem}{Theorem}
\newtheorem{lemma}{Lemma}

\theoremstyle{definition}
\newtheorem{definition}{Definition}

\newtheorem{corollary}{Corollary}

\newtheorem{remark}{Remark}

\begin{document}

\title{Generalized Lexicographic Products \\ and the Index Coding Capacity}
\author{
Fatemeh Arbabjolfaei and Young-Han Kim \\
Department of Electrical and Computer Engineering\\
University of California, San Diego\\
Email: \{farbabjo, yhk\}@ucsd.edu
}
\date{}
\maketitle

\begin{abstract}
The index coding problem studies the fundamental limit on broadcasting multiple messages
to their respective receivers with different sets of side information that are represented by a directed graph.
The generalized lexicographic product structure in the side information graph is introduced as a natural condition
under which the corresponding index coding problem can be decomposed into multiple interacting subproblems.
The capacity region is characterized in terms of the subproblem capacity regions
combined in the same product structure.
The proof is based on dual uses of random coding---one for a new multiletter characterization
of the capacity region of a general index coding problem via joint typicality decoding 
and the other for a construction of a new multiletter code of matching rates from a single-letter code
via joint typicality encoding. Several special cases are discussed that recover and strengthen known structural
properties of the index coding capacity region.
\end{abstract}

\section{Introduction}

\begin{figure}[t]
\centering
\small
\psfrag{x}[b]{$x_1,\ldots,x_n$}
\psfrag{m1}[b]{$y$}
\psfrag{ch}[c]{Channel}
\psfrag{e1}[c]{Encoder}
\psfrag{d1}[c]{Decoder $1$}
\psfrag{d0}[c]{Decoder $2$}
\psfrag{d3}[c]{Decoder $n$}
\psfrag{xh1}[b]{$x_1$}
\psfrag{xh0}[b]{$x_2$}
\psfrag{xh3}[b]{$x_n$}
\psfrag{a1}[b]{$x(A_1)$}
\psfrag{a2}[b]{$x(A_2)$}
\psfrag{a3}[b]{$x(A_n)$}
\includegraphics[scale=0.4]{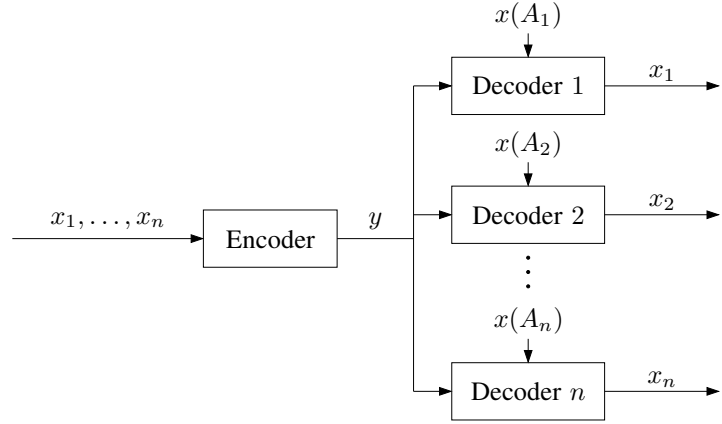}%
\caption{The index coding problem.}
\label{fig:index}
\end{figure}

\begin{figure}[t]
\centering
\small
\psfrag{1}[cb]{1}
\psfrag{2}[rc]{2}
\psfrag{3}[lc]{3}
\psfrag{4}{4}
\includegraphics[scale=0.4]{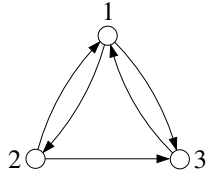}
\caption{The graph representation for the index coding problem with $A_1 = \{2,3\}, A_2 = \{1\}$, and $A_3 = \{1,2\}$.}
\label{fig:3-message}
\end{figure}

Index coding is a canonical problem in network information theory, in which a server broadcasts a tuple of $n$ messages $x^n = (x_1, \ldots, x_n)$, $x_i \in \{0,1\}^{t_i}$, to $n$ receivers by transmitting the fewest number of bits possible over a noiseless broadcast channel (see Fig.~\ref{fig:index}).
Receiver $i \in [n] := \{1,2, \ldots, n\}$ is interested in message $x_i$ and has a set of other messages $x(A_i) := (x_j, j \in A_i), A_i \subseteq [n] \setminus \{i\}$, as \emph{side information}. The \emph{side information sets} $A_1, \ldots, A_n$ are known to all communicating parties.
We represent the side information sets compactly by a sequence $(i|A_i), i \in [n]$.
For example, the 3-message index coding problem with $A_1 = \{2,3\}, A_2 = \{1\}$, and $A_3 = \{1,2\}$ is represented as
$(1|2,3), (2|1), (3|1,2)$.
Each index coding problem can be also uniquely specified by a (finite, simple) directed graph with $n$ vertices, referred to as the
{\em side information graph}.
Each vertex of the side information graph $G=(V,E)$ corresponds to a receiver (and its desired message) and there is a directed edge $j \to i$ if and only if (iff) receiver $i$ knows message $x_j$ as side information, i.e., $j \in A_i$ (see Fig.~\ref{fig:3-message}).
Throughout the paper, we identify an instance of the index coding problem with its side information graph 
$G$ and often write ``index coding problem $G$.''


We formulate the index coding problem more precisely by a $(t_1, \ldots, t_n, r)$ {\em index code} that consists of
\begin{itemize}
\item an encoder $\phi: \prod_{j \in [n]} \{0,1\}^{t_j} \to \{0,1\}^r$ that maps the message $n$-tuple $x^n$
to an $r$-bit sequence $y$ and
\item $n$ decoders, where the decoder at receiver $i \in [n]$,  $\psi_i: \{0,1\}^r \times \prod_{j \in A_i} \{0,1\}^{t_j} \to \{0,1\}^{t_i}$, maps the received sequence $y$ and the side information $x(A_i)$ back to $x_i$. 
\end{itemize}
Thus, for every $x^n \in \prod_{j \in [n]} \{0,1\}^{t_j}$,
\begin{equation} \label{eq:decoding-cond}
\psi_i(\phi(x^n), x(A_i)) = x_i, \quad i \in [n].
\end{equation}
Sometimes a $(t_1,\ldots,t_n,r)$ code will be written in short as a $(\tv,r)$ code and
a $(t, \ldots, t, r)$ code will be written in short as a $(t,r)$ code.

A rate tuple $\Rv = (R_1,\ldots,R_n)$ is said to be \emph{achievable} for the index coding problem $G$
if there exists a $(\tv, r)$ index code such that 
\[
R_i \le \frac{t_i}{r}, \quad i \in [n],
\]
or equivalently, in vector notation,
\[
\Rv \le \frac{\tv}{r}.
\]
Here and henceforth, we write $\av \le \bv$ for vectors $\av = (a_1,\ldots,a_n)$ and $\bv = (b_1,\ldots,b_n)$ of the same length $n$ iff $a_i \le b_i$, $i \in [n]$. 
The \emph{capacity region} $\Cr(G)$ 
of the index coding problem $G$ is defined as the closure of the set of achievable rate tuples.
The {\em symmetric capacity} (or the {\em capacity} in short) of the index coding problem $G$ is defined as
\[ 
C_\mathrm{sym}(G) = \max \{R \suchthat (R, \ldots, R) \in \Cr(G)\}.
\]
The reciprocal of the symmetric capacity, $\b(G) = 1/C_\mathrm{sym}(G)$, is referred to as the {\em broadcast rate}.

\begin{remark}
\label{rmk:epsilon-error}
The achievability of a given rate tuple can be defined alternatively by relaxing the decoding condition in~\eqref{eq:decoding-cond} as
\[
\lim_{r \to \infty} \P\{\psi_i(\phi(X^n), X(A_i)) \ne X_i,~i \in [n]\} = 0,
\]
where $X_1,\ldots,X_n$ are distributed independently and uniformly at random.
The corresponding \emph{vanishing-error capacity region}
can be shown  \cite{Willems1990} (see also \cite[Problem~8.11]{El-Gamal--Kim2011}) 
to be identical to the \emph{zero-error capacity region} defined earlier, which holds in general for any single-sender network.
This identity was also established in \cite{Chan--Grant2010, Langberg--Effros2011} 
in the context of index coding and single-sender network coding.
\end{remark}

The index coding problem was introduced by Birk and Kol~\cite{Birk--Kol1998} in 1998 and has been studied 
extensively over the past two decades. We refer the reader to the dissertations of El~Rouayheb \cite{El-Rouayheb2009},
Blasiak \cite{Blasiak2013}, and the first author \cite{Arbabjolfaei2017}, a survey article by Byrne and Calderini~\cite{Byrne--Calderini2018}, and a recent monograph by the authors~\cite{Arbabjolfaei--Kim2018}.
The main information-theoretic question in studying the index coding problem is 
to characterize the capacity region in a computable expression.
There are several inner and outer bounds on the capacity region that are tight for several interesting special cases, 
but the capacity region of a general $n$-message index coding problem is open (that is, no computable characterization is known).
So far the capacity region has been characterized for all index coding problems with $n \le 5$ messages
\cite{Arbabjolfaei--Bandemer--Kim--Sasoglu--Wang2013}.
For $n \ge 6$, the capacity region is not known in general.

\begin{figure}[t]
\begin{center}
\subfigure[]{
\small
\psfrag{1}[b]{}
\psfrag{2}[r]{}
\psfrag{3}[l]{}
\psfrag{4}[b]{}
\psfrag{5}[r]{}
\psfrag{6}[l]{}
\psfrag{g1}[t]{$G_1$}
\psfrag{g2}[t]{$G_2$}
\includegraphics[scale=0.4]{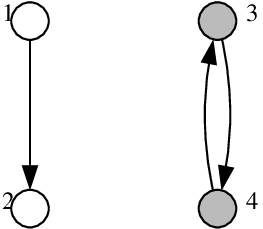}
}
\qquad \qquad
\subfigure[]{
\small
\psfrag{1}[b]{}
\psfrag{2}[r]{}
\psfrag{3}[l]{}
\psfrag{4}[b]{}
\psfrag{5}[r]{}
\psfrag{6}[l]{}
\psfrag{g1}[t]{$G_1$}
\psfrag{g2}[t]{$G_2$}
\includegraphics[scale=0.4]{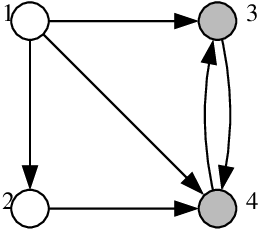}
}
\qquad \qquad
\subfigure[]{
\small
\psfrag{1}[b]{}
\psfrag{2}[r]{}
\psfrag{3}[l]{}
\psfrag{4}[b]{}
\psfrag{5}[r]{}
\psfrag{6}[l]{}
\psfrag{g1}[t]{$G_1$}
\psfrag{g2}[t]{$G_2$}
\includegraphics[scale=0.4]{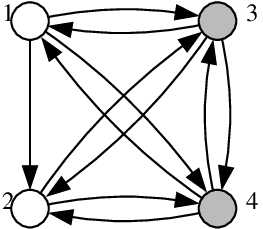}
}
\end{center}
\caption{Side information graphs with (a) no interaction, (b) one-way interaction, and (c) complete two-way interaction among the two parts (white and gray).}
\label{fig:no-oneway-complete-interact}
\end{figure}

For some cases, however, the side information can be decomposed into subgraphs with some connectivity (interaction) pattern, and this structure can be used to characterize the capacity region in terms of those of the subproblems. 
Consider the three side information graphs illustrated in Fig.~\ref{fig:no-oneway-complete-interact},
in which each graph has two parts and the interaction between them
is none, one-way, and complete two-way. These \emph{union} structures were investigated earlier in
\cite{Blasiak--Kleinberg--Lubetzky2011,Tahmasbi--Shahrasbi--Gohari2015,Arbabjolfaei--Kim2015a},
and it was shown that the capacity region of a given index coding problem is characterized as the ``sum''
of the subproblem capacity regions for the first two cases \cite{Tahmasbi--Shahrasbi--Gohari2015,Arbabjolfaei--Kim2015a},
and as the ``maximum'' of the subproblem capacity regions for the third case \cite{Arbabjolfaei--Kim2015a};
see Sections~\ref{subsec:no-interact} through~\ref{subsec:twoway} for details. 

\begin{figure}[b]
\centering
\subfigure[]{
\small
\psfrag{1}[b]{}
\psfrag{2}[r]{}
\psfrag{3}[l]{}
\psfrag{4}[b]{}
\psfrag{5}[r]{}
\psfrag{6}[l]{}
\raisebox{0.5em}{\includegraphics[scale=0.4]{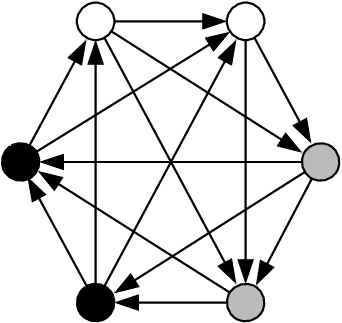}}
}
\qquad \qquad
\subfigure[]{
\small
\psfrag{1}[b]{}
\psfrag{2}[r]{}
\psfrag{3}[l]{}
\psfrag{4}[b]{}
\psfrag{5}[r]{}
\psfrag{6}[l]{}
\raisebox{0.15\height}{\includegraphics[scale=0.4]{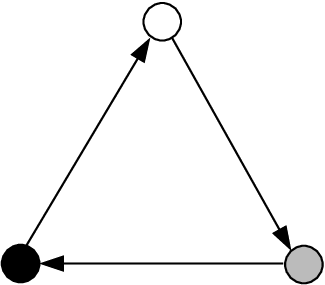}}
}
\qquad \qquad
\subfigure[]{
\small
\psfrag{1}[b]{}
\psfrag{2}[r]{}
\psfrag{3}[l]{}
\psfrag{4}[b]{}
\psfrag{5}[r]{}
\psfrag{6}[l]{}
\raisebox{1.6\height}{\includegraphics[scale=0.4]{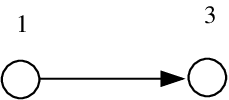}}
}
\caption{(a) The lexicographic product $G_0 \lexprod G_1$ of (b) the 3-vertex graph $G_0$ and (c) the 2-vertex graph $G_1$.}
\label{fig:lexprod}
\end{figure}

As another example, consider the side information graph in Fig.~\ref{fig:lexprod}(a), which can be 
generated by replacing each vertex of the graph in Fig.~\ref{fig:lexprod}(b) by the graph in Fig.~\ref{fig:lexprod}(c).
This \emph{lexicographic product} structure
was investigated in \cite{Blasiak--Kleinberg--Lubetzky2011}. 
Although the capacity in this case
was not characterized in terms of those of the subproblems, 
a ``product'' of the capacities of the subproblems was shown to be a nontrivial lower bound on the capacity, 
and this lower bound was utilized to establish a strong separation result between different capacity 
bounds~\cite{Blasiak--Kleinberg--Lubetzky2011}.

%


This paper identifies the \emph{generalized lexicographic product} structure as
a natural decomposition of the side information graph into subgraphs, which includes
the aforementioned 
union and product structures studied 
in \cite{Blasiak--Kleinberg--Lubetzky2011,Tahmasbi--Shahrasbi--Gohari2015,Arbabjolfaei--Kim2015a} as simple 
special cases.
The main contribution, presented in Theorem~\ref{thm:cap-generallexprod} in the next section,
shows that the capacity region of a generalized lexicographic product
has a natural lexicographic product structure itself, and
can be characterized in terms of the subgraph capacity regions
as well as the capacity region of the connectivity graph.
Although this generalized lexicographic product structure is rather special, 
its relaxation in Corollary~\ref{coro:noncritical} provides straightforward inner and outer bounds
on the capacity region for general side information graphs. 

%
%

The proof of Theorem~\ref{thm:cap-generallexprod} uses standard Shannon-theoretic arguments. 
The main challenge is the proof of the converse, which relies on two key ideas.
The first idea, Theorem~\ref{thm:capacity-it}, is a construction of an index code based on Shannon's \emph{random coding}
and \emph{joint typicality decoding}, the achievable rate region of which is characterized as a multiletter expression
by the \emph{packing lemma}.
The second idea, Lemma~\ref{lem:mappings}, is a construction of a new multiletter index code with relaxed decoding 
conditions from 
a single-letter index code,
which is based on 
random coding and joint typicality \emph{encoding}. The achievable rate region of this code is characterized by the \emph{covering lemma}.
The converse proof matches the corresponding rate regions
from the two ideas carefully to establish the desired structure of the capacity region.

The rest of the paper is organized as follows.
Section~\ref{sec:main_result} introduces the generalized lexicographic product of graphs and presents the capacity region 
of a generalized lexicographic product in terms of those of the subgraphs (Theorem~\ref{thm:cap-generallexprod}).
Section~\ref{sec:examples} presents several examples and special cases of
Theorem~\ref{thm:cap-generallexprod} and its relaxation (Corollary~\ref{coro:noncritical}).
Section~\ref{sec:random-coding} establishes the Shannon-theoretic multiletter characterization of the capacity region,
which may be of independent interest.
Section~\ref{sec:proof} presents the proof of Theorem~\ref{thm:cap-generallexprod}. 
Section \ref{sec:conclusion} concludes the paper with a few remarks on applications of the main result. 
Technical proofs used in the proof of the converse are relegated to the Appendices.

\section{Main Result}
\label{sec:main_result}

In this section, we first define the generalized lexicographic product of graphs and then state the main theorem of the paper.

\subsection{Generalized Lexicographic Product of Graphs}

Consider the following graph product, first considered by Schwenk \cite{Schwenk1974} in the context of spectral graph theory.

\begin{definition}[Generalized lexicographic product \cite{Schwenk1974, Godsil--McKay1978}]
Let $G_0 = (V(G_0),E(G_0))$ be a directed graph on $m$ vertices and let $G_i=(V(G_i),E(G_i))$, $i \in [m]$, be directed graphs on disjoint sets of vertices, i.e., $V(G_i) \cap V(G_j) = \emptyset$, $i \ne j$.
The \emph{generalized lexicographic product} $G = G_0 \lexprod (G_1, \ldots, G_m)$ is defined by the set of vertices $V(G) = \cup_{i \in [m]} V(G_i)$ and the set of edges $E(G)$ consisting of directed edges $(i,j)$ such that
\[
i,j \in V(G_k) ~\text{for some}~ k ~\text{and}~ (i,j) \in E(G_k)
\] 
\[
\text{or}
\]
\[
i \in V(G_k), j \in V(G_l) ~\text{for some}~ k \ne l ~\text{and}~ (k,l) \in E(G_0).
\]
\end{definition}

\begin{figure}[h]
\begin{center}
\subfigure[]{
\small
\psfrag{1}[bc]{1}
\psfrag{2}[bc]{2}
\psfrag{3}[l]{3}
\psfrag{4}[r]{4}
\psfrag{5}[r]{5}
\psfrag{6}[r]{6}
\includegraphics[scale=0.4]{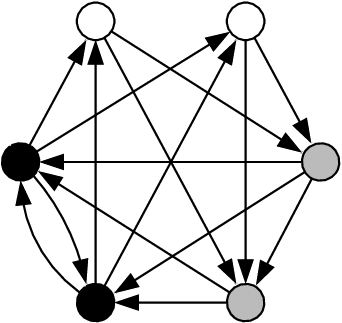}
}
\qquad
\subfigure[]{
\small
\psfrag{1}[b]{1}
\psfrag{2}[l]{2}
\psfrag{3}[r]{3}
\includegraphics[scale=0.4]{figs/3-node-triangle2.eps}
}
\\
\subfigure[]{
\small
\psfrag{1}[c]{}
\psfrag{3}[r]{}
\includegraphics[scale=0.4]{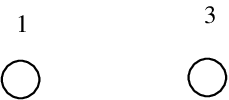}
}
\qquad
\subfigure[]{
\small
\psfrag{1}[c]{}
\psfrag{3}[c]{}
\includegraphics[scale=0.4]{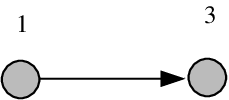}
}
\qquad
\subfigure[]{
\small
\psfrag{1}[c]{}
\psfrag{3}[r]{}
\includegraphics[scale=0.4]{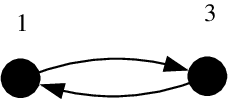}
}
\end{center}
\caption{(a) A 6-vertex graph that is the generalized lexicographic product $G_0 \lexprod (G_1, G_2, G_3)$, (b) the 3-vertex graph $G_0$, (c) the 2-vertex graph $G_1$, (d) the 2-vertex graph $G_2$, and (e) the 2-vertex graph $G_3$.}
\label{fig:general-lexprod}
\end{figure}

In other words, vertex $i \in V(G_0)$ is replaced by a copy of $G_i$ and every vertex in the copy of $G_k$ is connected to every vertex in the copy of $G_l$ according to $E(G_0)$; see Fig.~\ref{fig:general-lexprod} for an illustration.

\begin{remark}
This notion of generalized lexicographic product extends that of lexicographic product $G_0 \lexprod G_1$\cite{Hammack--Imrich--Klavzar2011,Scheinerman--Ullman2011}, which is a graph with vertex set $V(G_0) \times V(G_1)$ and $(i_1,i_2)$ is connected to $(j_1,j_2)$ iff
\[
(i_1,j_1) \in E(G_0) \quad \text{or} \quad \left(i_1 = j_1 \text{ and } (i_2,j_2) \in E(G_1) \right).
\]
By relabeling the vertices, $G_0 \lexprod G_1 = G_0 \lexprod (G_1^{(1)}, \ldots, G_1^{(m)})$, where $G_1^{(1)}, \ldots, G_1^{(m)}$ are copies of $G_1$ over disjoint vertex sets. 
\end{remark}

\begin{remark}
\label{rmk:verify-gen-lex-prod}
To verify whether a graph $G = ([n],E)$ is a generalized lexicographic product of smaller graphs, it suffices to go over all subsets of vertices $S \subseteq [n]$ with $2 \le |S| \le n-1$ and check if the vertices in $S$ have the same adjacency pattern with respect to all the vertices in $[n] \setminus S$.
\end{remark}

\subsection{Main Result}

The main contribution of the paper is the following characterization of the capacity region of the index coding problem $G_0 \lexprod (G_1, \ldots, G_m)$ in terms of the capacity regions of smaller problems $G_0, G_1, \ldots, G_m$.

\begin{theorem}
\label{thm:cap-generallexprod}
Let $G_0=([m],E)$ be the side information graph of an index coding problem with $m$ messages and capacity region $\Cr_0$. 
Let $G_1, \ldots, G_m$ be the side information graphs of $m$ index coding problems with capacity regions $\Cr_1, \ldots, \Cr_m$, respectively. 
Then the capacity region of the index coding problem with side information graph $G = G_0 \lexprod (G_1, \ldots, G_m)$ is \begin{align}
\label{eq:cap-generallexprod}
\Cr(G) &= \Cr_0 \lexprod(\Cr_1,\ldots,\Cr_m) \notag \\
&:= \bigl\{ (\rho_1 \Rv_1, \ldots, \rho_m \Rv_m) \suchthat \rhov \in \Cr_0,~\Rv_i \in \Cr_i, i \in [m] \bigr\}
\end{align}
and its broadcast rate is 
\begin{align}
\label{eq:beta-generallexprod}
\b(G) = \min_{R \suchthat (R,\ldots,R) \in \Cr(G)} \frac{1}{R}.
\end{align}
\end{theorem}

\begin{remark}
\label{rmk:cap-generallexprod}
Since $\Cr_0,\Cr_1,\ldots,\Cr_m$ are compact, so is the RHS of \eqref{eq:cap-generallexprod}.
\end{remark}

\begin{remark}
\label{rmk:polytope}
If $\Cr_0, \Cr_1, \ldots, \Cr_m$ are polytopes of the form $\Cr_i = \{\Rv: T_i\Rv \le \mathbf{1}=(1,\ldots,1)^T\}$, $i=0,1,\ldots,m$, then $\Cr$ is also a polytope characterized by Fourier--Motzkin elimination of $m$ variables $\rhov = (\rho_1,\ldots,\rho_m)$ from the linear inequalities
\begin{align*}
T_0 \rhov &\le \mathbf{1}, \\
T_i\Rv_i &\le \rho_i\mathbf{1}, \quad i \in [m].
\end{align*}
\end{remark}

\begin{remark}
Theorem~\ref{thm:cap-generallexprod} can be specialized to the broadcast rate of $G = G_0 \lexprod (G_1, \ldots, G_m)$.
If $\betav = (\b(G_1),\ldots, \b(G_m))$, then
\[
\b(G) = \frac{1}{C_0(\betav)} \le \max_{i \in [m]} \b(G_0)\b(G_i),
\]
where $C_0(\betav) = \max \{R \suchthat R \betav \in \Cr_0\}$.
\end{remark}

The following sandwich argument extends the application of Theorem~\ref{thm:cap-generallexprod} beyond index coding instances with side information graph in the form of a generalized lexicographic product.

\begin{corollary}
\label{coro:noncritical}
For $i=0,1,\ldots,m$, let $G'_i$ and $G''_i$ be side information graphs of index coding problems with capacity regions $\Cr_i'$ and $\Cr_i''$, respectively, such that $V(G'_i) = V(G''_i)$ and $E(G'_i) \subseteq E(G''_i)$.
Suppose that $|V(G'_0)| = |V(G''_0)| = m$ and let 
\[
G' = G'_0 \lexprod (G'_1, \ldots, G'_m)
\]
and
\[
G'' = G''_0 \lexprod (G''_1, \ldots, G''_m).
\]
Then the capacity region of any index coding problem $G$ such that
\[
V(G) = V(G') = V(G'')
\]
and
\[
E(G') \subseteq E(G) \subseteq E(G'')
\]
is bounded as 
\begin{align*}
\Cr_0' \lexprod(\Cr_1',\ldots,\Cr_m') 
= \Cr(G') 
\subseteq \Cr(G) 
\subseteq \Cr(G'') = \Cr_0'' \lexprod(\Cr_1'',\ldots,\Cr_m'').
\end{align*}
In particular, if $\Cr_i' = \Cr_i'' = \Cr_i$, $i = 0, 1,\ldots,m$, then 
\begin{align*}
\Cr(G) = \Cr(G') = \Cr(G'') = \Cr_0 \lexprod(\Cr_1,\ldots,\Cr_m).
\end{align*}
\end{corollary}


\begin{remark}
\label{rmk:verify-beyond-gen-lex-prod}
For any side information graph $G$, the bounding graphs $G'$ and $G''$ can be easily constructed by
considering any vertex subset $S$, say $[k]$, with $2 \le |S| = k \le n-1$,
and taking the intersection and union of the neighbors from/to $S$ to/from $[n]\setminus S$, respectively.
Now that the adjacency pattern is the same for all vertices in $S$, we can identify
$G_0'$ and $G_0''$ by replacing $G|_S$ with a single vertex and keeping the other vertices.
The resulting $G'$ and $G''$ are generalized lexicographic products of $n-k+1$ graphs.
\end{remark}

\begin{remark}
\label{rmk:critical}
An index coding problem is said to be \emph{critical} if removal of any of the edges of its side information graph 
strictly reduces the capacity region \cite{Tahmasbi--Shahrasbi--Gohari2015, Arbabjolfaei--Kim2015a}.
Note that in Corollary~\ref{coro:noncritical}, $\Cr'_i = \Cr''_i$, $i = 0,1,\ldots,m,$ implies that the index coding problem $G$ is not critical, as those edges of the side information graph $G$ that are not in $G'$ 
can be removed from $G$ without reducing the capacity region.
Thus, Corollary~\ref{coro:noncritical} provides a necessary condition for criticality of a side information graph (see \cite{Arbabjolfaei--Kim2018} for other necessary conditions).
%
\end{remark}

\section{Examples}
\label{sec:examples}

\subsection{No Interaction Between Partitions}
\label{subsec:no-interact}

\begin{figure}[t]
\begin{center}
\subfigure[]{
\psfrag{1}{}
\psfrag{3}{}
\includegraphics[scale=0.4]{figs/G0no-interact.eps}
}
\qquad
\subfigure[]{
\psfrag{1}{}
\psfrag{3}{}
\includegraphics[scale=0.4]{figs/G0one-way.eps}
}
\qquad
\subfigure[]{
\psfrag{1}{}
\psfrag{3}{}
\includegraphics[scale=0.4]{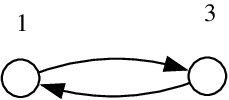}
}
\end{center}
\caption{(a) A 2-vertex graph with no edge, (b) a 2-vertex graph with one edge, and (c) a 2-vertex graph with two edges.}
\label{fig:G0}
\end{figure}

Consider the side information graph $G$ depicted in Fig.~\ref{fig:no-oneway-complete-interact}(a), which has two noninteracting parts $G_1$ and $G_2$, i.e., there is no edge between $G_1$ and $G_2$.
Then $G$ can be viewed as $G_0 \lexprod (G_1, G_2)$, where $G_0$ is the two-vertex graph in Fig.~\ref{fig:G0}(a).
Since the capacity region of $G_0$ is $\{(R_1,R_2) \suchthat R_1 + R_2 \le 1\}$, by Theorem~\ref{thm:cap-generallexprod},
\begin{align}
\label{eq:cap-no-interact}
\Cr(G) = \bigl\{ (\rho \Rv_1, (1-\rho) \Rv_2) \suchthat \Rv_1 \in \Cr(G_1), \Rv_2 \in \Cr(G_2), \rho \in [0,1] \bigr\}.
\end{align}
Moreover, the maximum symmetric rate in $\Cr(G)$ is attained when $\rho/\b(G_1) = (1-\rho)/\b(G_2)$, or equivalently, $\rho = \b(G_1)/(\b(G_1)+\b(G_2))$, which implies
\begin{align}
\label{eq:beta-no-interact}
\b(G) = \b(G_1) + \b(G_2).
\end{align}

More generally, consider a side information graph $G$ that consists of $m$ vertex-induced subgraphs $G_1, \ldots, G_m$ with no edges among them.
Then $G$ can be viewed as $G_0 \lexprod (G_1, \ldots, G_m)$, where $G_0$ is a graph with $m$ vertices and no edge.
By Theorem~\ref{thm:cap-generallexprod} (or by applying \eqref{eq:cap-no-interact} and \eqref{eq:beta-no-interact} inductively),
\begin{align*}
\Cr(G) = \biggl\{ (\rho_1 \Rv_1, \ldots,  \rho_m \Rv_m) \suchthat \Rv_i \in \Cr(G_i), i \in [m], \sum_{i \in [m]} \rho_i \le 1 \biggr\}
\end{align*}
and
\begin{align*}
\b(G) = \sum_{i \in [m]} \b(G_i).
\end{align*}
In other words, when $G$ is partitioned into noninteracting parts $G_1, \ldots, G_m$, the capacity region of $G$ is achieved  by \emph{time division} among the optimal coding schemes for subproblems $G_1, \ldots, G_m$ \cite{Tahmasbi--Shahrasbi--Gohari2015}.

\subsection{One-way Interaction Between Partitions}
\label{subsec:oneway}

Consider the side information graph $G$ depicted in Fig.~\ref{fig:no-oneway-complete-interact}(b), which has one-way interaction between its two parts $G_1$ and $G_2$, i.e., there is no edge from $G_2$ to $G_1$.
Let $G'_0$ and $G''_0$ be the two graphs on two vertices as depicted in Figs.~\ref{fig:G0}(a) and (b), respectively.
Then, $\Cr(G'_0) = \Cr(G''_0) = \{(R_1,R_2) \suchthat R_1 + R_2 \le 1\}$ and $E(G'_0 \lexprod (G_1,G_2)) \subseteq E(G) \subseteq E(G''_0 \lexprod (G_1,G_2))$.
Thus, by Corollary \ref{coro:noncritical},
\begin{align*}
\Cr(G) = \bigl\{ (\rho \Rv_1, (1-\rho) \Rv_2) \suchthat \Rv_1 \in \Cr(G_1), \Rv_2 \in \Cr(G_2), \rho \in [0,1] \bigr\}
\end{align*}
and
\begin{align*}
\b(G) = \b(G_1) + \b(G_2).
\end{align*}

More generally, suppose that the graph $G$ consists of $m$ vertex-induced subgraphs $G_1, \ldots, G_m$ such that there exists no edge from $G_j$ to $G_i$ for $i < j$.
Let $G'_0$ and $G''_0$ be directed graphs with $m$ vertices such that $E(G'_0) = \emptyset$ and $E(G''_0) = \{(i,j)\suchthat i < j\}$.
Note that $\Cr(G'_0) = \Cr(G''_0) = \{(R_1,\ldots,R_m) \suchthat \sum_{i \in [m]} R_i \le 1\}$.
Since $E(G'_0 \lexprod (G_1, \ldots, G_m)) \subseteq E(G) \subseteq E(G''_0 \lexprod (G_1, \ldots, G_m))$, by Corollary~\ref{coro:noncritical},
\begin{align*}
\Cr(G) = \biggl\{ (\rho_1 \Rv_1, \ldots,  \rho_m \Rv_m) \suchthat \Rv_i \in \Cr(G_i), i \in [m], \sum_{i \in [m]} \rho_i \le 1 \biggr\}
\end{align*}
and
\begin{align*}
\b(G) = \sum_{i \in [m]} \b(G_i).
\end{align*}
In words, the capacity region of a graph with one-way interaction among its parts is no larger than the capacity region of a graph with noninteracting parts. Thus the edges connecting the parts $G_1, \ldots, G_m$ in one way,
or equivalently by the Farkas lemma~\cite[Th.~2.2]{Achim--Kern1992}, the edges that are not on a directed cycle
can be removed without affecting the capacity region (cf. Remark~\ref{rmk:critical}) and the graph is not critical \cite{Tahmasbi--Shahrasbi--Gohari2015}.

\subsection{Complete Two-way Interaction Between Partitions}
\label{subsec:twoway}

Consider the side information graph $G$ in Fig.~\ref{fig:no-oneway-complete-interact}(c).
Since there are two-way edges between every vertex in $G_1$ and every vertex in $G_2$, $G$ can be written as $G_0 \lexprod (G_1,G_2)$, where $G_0$ is the complete graph with two vertices depicted in Fig.~\ref{fig:G0}(c).
More generally, suppose that $G_0$ is a complete graph with $m$ vertices.
Then its capacity region is characterized as $\Cr(G_0) = \{(R_1,\ldots,R_m) \suchthat R_i \le 1, i \in [m]\}$.
Thus, by Theorem~\ref{thm:cap-generallexprod}, the capacity region of $G = G_0 \lexprod (G_1, \ldots, G_m)$ is
\begin{align}
\label{eq:cap-G0-complete}
\Cr(G) = \bigl\{ (\Rv_1, \ldots, \Rv_m) \suchthat \Rv_i \in \Cr(G_i), i \in [m] \bigr\}.
\end{align}
Moreover, \eqref{eq:cap-G0-complete} implies
\begin{align*}
\max \{R: (R, \ldots, R) \in \Cr(G)\} = \min_{i \in [m]} \max \{R: (R, \ldots, R) \in \Cr_i\} = \min_{i \in [m]} \frac{1}{\b(G_i)}
\end{align*}
and thus
\begin{align*}
\b(G) = \max_{i \in [m]} \b(G_i).
\end{align*}
In words, the capacity region of a graph with complete two-way interaction among its parts  is achieved by simultaneously using the optimal coding schemes for individual parts \cite{Arbabjolfaei--Kim2015a}.

\subsection{Lexicographic Products}
\label{subsec:lex}

We revisit the side information graph $G$ in Fig.~\ref{fig:lexprod}(a), which is the lexicographic product
of the two graphs in~Fig.~\ref{fig:lexprod}(b) and Fig.~\ref{fig:lexprod}(c).
By Theorem~\ref{thm:cap-generallexprod}, the capacity region of problem $G_0 \lexprod G_1$ is
\begin{align*}
\Cr(G) = \bigl\{ (\rho_1\Rv_1, \ldots, \rho_m\Rv_m) \suchthat \rhov \in \Cr(G_0),~\Rv_i \in \Cr(G_1), i \in [m] \bigr\},
\end{align*}
which implies
\begin{align}
\label{eq:beta-lexprod}
\beta(G_0 \lexprod G_1) = \beta(G_0) \beta(G_1).
\end{align}
In words, the broadcast rate is multiplicative under the lexicographic product of index coding side information graphs.
We note that one direction ($\le$) in \eqref{eq:beta-lexprod} was established earlier in \cite{Blasiak--Kleinberg--Lubetzky2011}.


\subsection{Beyond Generalized Lexicographic Products}
\label{subsec:beyond-glex}

\begin{figure}[t]
\begin{center}
\subfigure[]{
\small
\psfrag{1}[b]{1}
\psfrag{2}[l]{2}
\psfrag{3}[l]{3}
\psfrag{4}[r]{4}
\psfrag{5}[r]{5}
\includegraphics[scale=0.4]{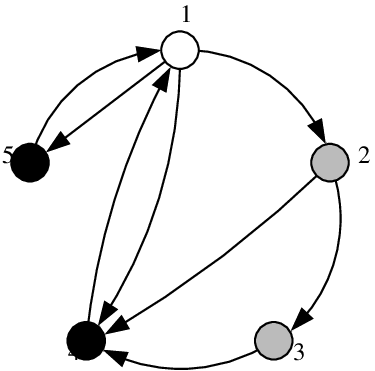}%
}
\qquad \qquad
\subfigure[]{
\small
\psfrag{1}[b]{1}
\psfrag{2}[l]{2}
\psfrag{3}[l]{3}
\psfrag{4}[r]{4}
\psfrag{5}[r]{5}
\includegraphics[scale=0.4]{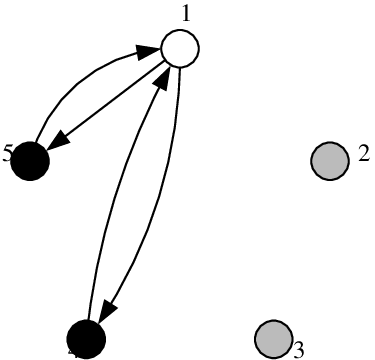}%
}
\qquad \qquad
\subfigure[]{
\small
\psfrag{1}[b]{1}
\psfrag{2}[l]{2}
\psfrag{3}[l]{3}
\psfrag{4}[r]{4}
\psfrag{5}[r]{5}
\includegraphics[scale=0.4]{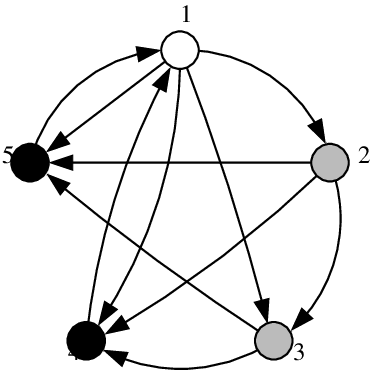}%
}
\end{center}
\caption{(a) The 5-vertex graph $G$ is sandwiched between (b) $G' = G'_0 \lexprod (G'_1,G'_2,G'_3)$ and (c) $G'' = G''_0 \lexprod (G''_1,G''_2,G''_3)$.}
\label{fig:noncritical-lexprod}
\end{figure}

\begin{figure}[t]
\begin{center}
\subfigure[]{
\small
\psfrag{1}[b]{a}
\psfrag{2}[r]{c}
\psfrag{3}[l]{b}
\includegraphics[scale=0.4]{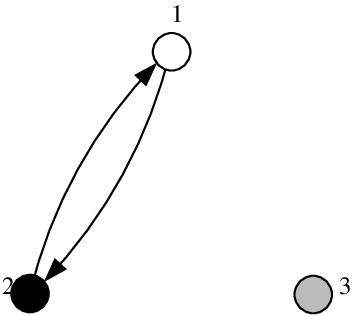}
}
\qquad \qquad
\subfigure[]{
\small
\psfrag{1}[b]{a}
\psfrag{2}[r]{c}
\psfrag{3}[l]{b}
\includegraphics[scale=0.4]{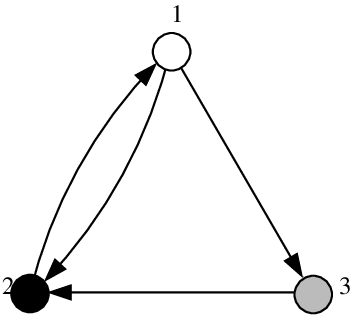}
}
\\[2em]
\subfigure[]{
\small
\psfrag{1}[b]{1}
\includegraphics[scale=0.4]{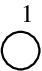}
}
\qquad\qquad
\subfigure[]{
\small
\psfrag{1}[b]{2}
\psfrag{3}[b]{3}
\includegraphics[scale=0.4]{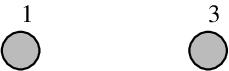}
}
\qquad \qquad
\subfigure[]{
\small
\psfrag{1}[b]{2}
\psfrag{3}[b]{3}
\includegraphics[scale=0.4]{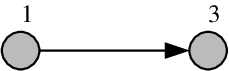}
}
\qquad \qquad
\subfigure[]{
\small
\psfrag{1}[b]{4}
\psfrag{3}[b]{5}
\includegraphics[scale=0.4]{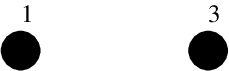}
}
\end{center}
\caption{(a) The 3-vertex graph $G'_0$, (b) the 3-vertex graph $G''_0$, (c) the 1-vertex graph $G'_1 = G''_1$, (d) the 2-vertex graph $G'_2$, (e) the 2-vertex graph $G''_2$, and (f) the 2-vertex graph $G'_3 = G''_3$.}
\label{fig:noncritical-lexprod-comp}
\end{figure}

In Section~\ref{subsec:oneway}, we have seen a simple application of Corollary~\ref{coro:noncritical}.
We now present a more substantial example.
Consider the side information graph depicted in Fig.~\ref{fig:noncritical-lexprod}(a), which cannot be viewed as the generalized lexicographic product of smaller graphs.
Let $G'$ and $G''$ be the graphs depicted in Figs.~\ref{fig:noncritical-lexprod}(b) and (c), respectively.
Since the graph $G$ satisfies
$V(G) = V(G') = V(G'')$
and
$E(G') \subseteq E(G) \subseteq E(G'')$,
its capacity region is sandwiched between the capacity regions $\Cr(G')$ and $\Cr(G'')$.
Now the graphs $G'$ and $G''$ are generalized lexicographic products of smaller
graphs as $G' = G'_0 \lexprod (G'_1,G'_2,G'_3)$ and $G'' = G''_0 \lexprod (G''_1,G''_2,G''_3)$, where  $G'_0$, $G''_0$, $G'_1=G''_1$, $G'_2$, $G''_2$, and $G'_3=G''_3$ are the graphs depicted in Fig.~\ref{fig:noncritical-lexprod-comp}.
Note that for each $i = 0,1,2,3$, $V(G'_i) = V(G''_i)$ and $E(G'_i) \subseteq E(G''_i)$.
Furthermore, the capacity regions of problems $G'_i$ and $G''_i$ can be shown to be identical as
\begin{align*}
\Cr_0 &= \Cr(G'_0) = \Cr(G''_0) = \{(\rho_a,\rho_b,\rho_c): \rho_a + \rho_b \le 1, \rho_b + \rho_c \le 1\}, \\
\Cr_1 &= \Cr(G'_1) = \Cr(G''_1) = \{R_1: R_1 \le 1\}, \\
\Cr_2 &= \Cr(G'_2) = \Cr(G''_2) = \{(R_2,R_3): R_2 + R_3 \le 1\}, \\
\Cr_3 &= \Cr(G'_3) = \Cr(G''_3) = \{(R_4,R_5): R_4 + R_5 \le 1\}.
\end{align*}


\noindent Hence, by Corollary~\ref{coro:noncritical}, the capacity region $\Cr(G)$ of index coding problem $G$ is equal to $\Cr(G') = \Cr(G'')$, which is the set of all rate tuples $(R_1,R_2,R_3,R_4,R_5)$ such that
\begin{align*}
R_1 &\le \rho_a, \\
R_2 + R_3 &\le \rho_b, \\
R_4 + R_5 &\le \rho_c
\end{align*}
for some $(\rho_a,\rho_b,\rho_c)$ such that $\rho_a+\rho_b \le 1$ and $\rho_b + \rho_c \le 1$.
By Remark~\ref{rmk:polytope}, this region simplifies to the set of $(R_1,\ldots, R_5)$ such that
\begin{align*}
R_1 + R_2 + R_3 &\le 1,\\
R_2 + R_3 + R_4 + R_5 &\le 1.
\end{align*}

\section{A Multiletter Characterization of the Index Coding Capacity Region}
\label{sec:random-coding}

Alon, Hassidim, Lubetzky, Stav, and Weinstein \cite{Alon--Hassidim--Lubetzky--Stav--Weinstein2008} established a graph-theoretic characterization of the broadcast rate as the limit
of multiletter expressions involving the chromatic number of the confusion graph \cite{Alon--Hassidim--Lubetzky--Stav--Weinstein2008, Gadouleau--Riis2011}. This characterization was later strengthened in \cite{Arbabjolfaei--Kim2016a} by replacing the chromatic number
with the \emph{fractional} chromatic number and also extended 
 to the capacity region.

In this section, we use Shannon's random coding idea \cite{Shannon1948} to establish the following information theoretic multiletter characterization of the capacity region  of the index coding problem.

\begin{theorem}
\label{thm:capacity-it}
The capacity region of the index coding problem $(i|A_i)$, $i \in [n]$, with side information graph $G$ is the closure of
\[
\bigcup_{r=1}^\infty \Cr_r(G),
\]
where $\Cr_r(G)$ is the set of all rate tuples $(R_1, \ldots, R_n)$ satisfying
\begin{align*} 
R_i & \le \frac{1}{r}I(U_i; V | U(A_i)), \quad i \in [n],
\end{align*}
for some pmf $p(u_1) \cdots p(u_n)$ and function
$f: \Uc_1\times\cdots\times\Uc_n \to \Vc$ that maps the $n$-tuple $(U_1, \ldots, U_n)$ to $V$ such that the cardinalities of the auxiliary random variables $U_1,\ldots,U_n,$ and $V$ are upper bounded by $2^r$.
\end{theorem}


Here $I(U_i; V | U(A_i))$ denotes the conditional mutual information \cite{Shannon1948,Cover--Thomas2006} between $U_i$ and $V$ given $U(A_i)$. Since $U_1,\ldots,U_n$ are mutually independent,
\[
I(U_i;V | U(A_i)) = I(U_i; V, U(A_i)), \quad i \in [n].
\]
In the following, we prove the theorem in two steps.

\subsection{Proof of Achievability}

We follow the standard arguments in the random coding proof of Shannon's channel coding theorem using the notion of typicality \cite{Shannon1948,Cover--Thomas2006,El-Gamal--Kim2011}.
Here and henceforth,
we define the set of $\e$-typical $k$-sequences $u^k = (u_1,\ldots,u_k)$ with respect to $U \sim p(u)$ for $\e \in (0,1)$ as
\begin{align*}
\aepk(U) = \{u^k: |\pi(u|u^k)-p(u)| \le \e p(u) ~\text{for all}~ u \in \Uc\},
\end{align*}
where
\begin{align*}
\pi(u|u^k) = \frac{|i: u_i = u|}{k}, \quad u \in \Uc,
\end{align*}
is the empirical pmf of $u^k$. Elementary properties of the typical set and typical sequences can be found in \cite{Orlitsky--Roche2001,El-Gamal--Kim2011}.

Now we prove the achievability of the rate tuples in $\Cr_r$ for each $r = 1,2,\ldots,$ based on random coding.
For simplicity of presentation, we assume throughout the proof that $kr R_i$ is an integer for every $i \in [n]$.

{\em Codebook generation.} Fix a pmf $p(u_1) \cdots p(u_n)$ and a function $v = f(u_1, \ldots, u_n)$ under the prescribed cardinality constraints.
For each $i \in [n]$, randomly and independently generate $2^{kr R_i}$ sequences $u_i^k(x_i)$, $x_i \in [2^{krR_i}]$, each according to $\prod_{j = 1}^k p_{U_i}(u_{ij})$.
These codewords constitute the codebook, which is shared among all communicating parties.

{\em Encoding.} To communicate the message tuple $(x_1,\ldots,x_n)$, we transmit $y = v^k(u_1^k(x_1),\ldots,u_n^k(x_n)) \in [2^{kr}]$, where $v_j = f(u_{1j}(x_1), \ldots, u_{nj}(x_n))$, $j \in [k]$.

\medskip

{\em Decoding.} We use \emph{joint typicality decoding} (see, for example, \cite[Sec.~3.1]{El-Gamal--Kim2011}).
Let $v^k$ be the received sequence and $u^k(x(J)) = (u_j^k(x_j), j \in J)$.
Decoder $i \in [n]$ declares that $\xh_i$ is sent if it is the unique message such that
\[
(u_i^k(\xh_i), u^k(x(A_i)), v^k) \in \aepk.
\]
Otherwise it declares an error.

{\em Analysis of the probability of error.}
By the symmetry of codebook generation, the probability of error averaged over the messages
and the random codebook generation satisfies
\[
\P(\Ec) = \P\{(X_1,\ldots,X_n) \ne (\Xh_1,\ldots,\Xh_n) \} = \P\{\Ec|(X_1,\ldots,X_n) = (1,\ldots,1)\}.
\]
Hence, we assume without loss of generality that $X_i = 1$, $i \in [n]$, is sent,
and suppress the condition $\{(X_1,\ldots,X_n) = (1,\ldots,1)\}$ in the subsequent probability expressions for brevity.
Note that decoder~$i$ makes an error iff one or more of the following events occur:
\begin{align*}
\mathcal{E}_{i1} & = \{(U_i^k(1), U^k((1,\ldots,1)), V^k) \not \in \aepk\}, \\
\mathcal{E}_{i2} & = \{(U_i^k(x_i), U^k((1,\ldots,1)), V^k) \in \aepk ~\text{for some}~ x_i \ne 1\}.
\end{align*}
Thus, by the union of events bound, the probability of error for decoder $i$ is upper bounded as
\begin{align*}
\P(\Ec_{i}) \le \P(\Ec_{i1}) + \P(\Ec_{i2}).
\end{align*}
By the law of large numbers, $\P(\Ec_{i1})$ tends to zero as $k \to \infty$.
If $x_i \not = 1$, $U_i^k(x_i)$ is independent of $V^k$ and $U^k(A_i)$.
Hence, by the packing lemma \cite[Lemma 3.1]{El-Gamal--Kim2011}, $\P(\Ec_{i2})$ tends to zero as $k \to \infty$ if
\begin{align} \label{eq:rate-constraint}
rR_i < I(U_i;V,U(A_i)) - \d(\e) = I(U_i;V|U(A_i)) - \d(\e),
\end{align}
where $\d(\e)$ tends to zero as $\e \to 0$ and the last identity follows since $U_i$ and $U(A_i)$ are independent.
Therefore, if the specified rate constraints in~\eqref{eq:rate-constraint} are satisfied simultaneously for all messages,
the probability of error $\P(\Ec)$ averaged over messages and codebooks
tends to zero as $k \to \infty$, and there must exist a sequence
of $(\ceil{krR_1},\ldots,\ceil{krR_n},kr)$ index codes such that the probability of error averaged over the messages
tends to zero as $k \to \infty$. Letting $\e \to 0$ shows that any rate tuple $(R_1,\ldots,R_n) \in \Cr_r$ is achievable
with vanishing probability of error.
By Remark \ref{rmk:epsilon-error}, this error probability can be made to be exactly zero without
sacrificing the rates and thus $\Cr_r$ is contained in the capacity region.
This completes the proof of achievability.


\subsection{Proof of the Converse}
We show that any achievable rate tuple $(R_1,\ldots, R_n)$ lies in some $\Cr_r$. First note that
for any $(t_1,\ldots,t_n,r)$ index code, 
\begin{align*}
H(X_i|Y,X(A_i)) &= 0, \quad i \in [n].
\end{align*}
Hence,
\[
rR_i \le t_i = H(X_i) = I(X_i;Y|X(A_i)), \quad i \in [n].
\]
By identifying $U_i = X_i$, $i \in [n]$, and $V = Y$, the cardinalities
of which are all upper bounded by $2^r$, we can conclude that
\[
R_i \le \frac{1}{r} I(U_i; V | U(A_i)), \quad i \in [n],
\]
for some $p(u_1)\cdots p(u_n)$ and $v = f(u_1,\ldots,u_n)$ such that
the cardinalities are bounded by $2^r$.
This completes the proof of the converse.

\section{Proof of Theorem~\ref{thm:cap-generallexprod}}
\label{sec:proof}

In this section, we use the information theoretic characterization of index coding capacity region in Theorem~\ref{thm:capacity-it}  to prove the main result of the paper.

\subsection{Proof of Achievability}

\begin{figure}[h]
\centering
\small
\psfrag{x}[b]{$y$}
\psfrag{e}[c]{Encoder $G_0$}
\psfrag{m1}[bc]{$\xv_1$}
\psfrag{m2}[bc]{$\xv_i$}
\psfrag{m3}[bc]{$\xv_m$}
\psfrag{s1}[bc]{$y_1$}
\psfrag{s2}[bc]{$y_i$}
\psfrag{s3}[bc]{$y_m$}
\psfrag{e1}[c]{Encoder $G_1$}
\psfrag{e2}[c]{Encoder $G_i$}
\psfrag{e3}[c]{Encoder $G_m$}
\includegraphics[scale=0.4]{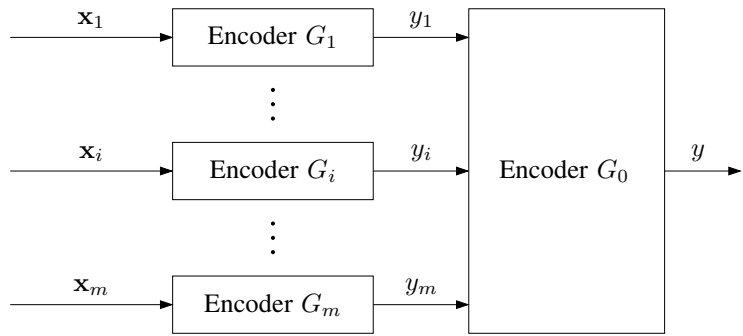}
\caption{Construction of an index code for index coding problem $G_0 \lexprod (G_1,\ldots,G_m)$ by concatenating the index codes for problems $G_1, \ldots ,G_m$ as the inner codes and the index code for problem $G_0$ as the outer code. The message tuple $\xv = (\xv_1, \ldots,\xv_m)$ is encoded by index codes for $G_1,\ldots,G_m$ part by part. The outputs $y_1,\ldots,y_m$ are then encoded by the index code for $G_0$.}
\label{fig:lexprod-ach}
\end{figure}

The proof of achievability extends the arguments in \cite{Blasiak--Kleinberg--Lubetzky2011}
and uses the simple construction of an index code for $G = G_0 \lexprod (G_1, \ldots, G_m)$ from index codes for subproblems as illustrated in Fig.~\ref{fig:lexprod-ach}.
To be more precise, consider any rate tuple $(\rho_1 \Rv_1, \ldots, \rho_m \Rv_m)$, where $\Rv_i \in \Cr_i$, $i \in [m]$, and $(\rho_1,\ldots,\rho_m) \in \Cr_0$.
Let $\e > 0$.
Then, by the definition of the capacity region, there exists a $(\ceil{(\rho_1-\e)r}, \ldots, \ceil{(\rho_m-\e)r} ,r)$ index code for problem $G_0$ for $r$ sufficiently large.
Also for each $i \in [m]$, there exists a $(\ceil{(\Rv_i-\e\mathbf{1})r_i} ,r_i)=(\tv_i,r_i)$ index code for problem $G_i$ for $r_i$ sufficiently large.
Let $r_i = \ceil{(\rho_i-\e)r}$, $i \in [m]$.
Then, by concatenating the $(\tv_i,r_i)$ index codes, $i \in [m]$, with $(r_1,\ldots,r_m,r)$ index code as shown in Fig.~\ref{fig:lexprod-ach}, we can construct a code for problem $G$.
The rate of message $i$ of this code is
\begin{align*}
\frac{\tv_i}{r} &= \frac{r_i}{r} \frac{\tv_i}{r_i} \\
&= \frac{\ceil{(\rho_i-\e)r}}{r} \frac{\ceil{(\Rv_i-\e\mathbf{1})r_i}}{r_i} \\
&\ge (\rho_i-\e)(\Rv_i-\e\mathbf{1}), \quad i \in [m].
\end{align*}
Letting $\e \to 0$ completes the proof.

\subsection{Proof of the Converse}
Our proof is inspired by the proof for the one-way interaction
in \cite{Tahmasbi--Shahrasbi--Gohari2015}, but significantly extends the arguments therein.
Let $A_j \subseteq V(G)$ denote the side information set of receiver~$j \in V(G)$ for the index coding problem $G$, and let $A'_i \subseteq [m]$
denote the side information set of receiver~$i \in [m]$ for the index coding problem $G_0$. In this notation, if $j \in V(G_i)$ for some $i \in [m]$, then
by the definition of generalized lexicographic product, the side information set of receiver~$j$ can be decomposed as
\begin{align} \label{eq:side-decomposition}
A_j = \bigl(A_j \cap V(G_i)\bigr) \cup \biggl( \bigcup_{l \in A'_i} V(G_l) \biggr),
\end{align}
where the first term denotes the side information from within the subproblem $G_i$ and the second term denotes the side information from other subproblems.
As in Fig.~\ref{fig:lexprod-ach}, we write $\xv_i$ for $(x_j: j \in V(G_i))$ and $\xv$ for $(\xv_1, \ldots, \xv_m)$. We also write
$\xv(A'_i)$ for $(x_j: j \in \cup_{l \in A'_i} V(G_l))$.

To prove the converse ($\Cr \subseteq \Cr_0 \lexprod (\Cr_1,\ldots,\Cr_m)$), for any $(\tv_1, \ldots, \tv_m, r)$ index code for $G = G_0 \lexprod (G_1,\ldots,G_m)$, we argue that the corresponding rate tuple can be factored as
\[
\frac{\tv_i}{r} = \frac{s_i}{r} \frac{\tv_i}{s_i}, \quad i \in [m],
\]
for some $(s_1,\ldots,s_m)$, so that
\begin{subequations} \label{eq:lex-proof-steps}
\begin{align}
\biggl(\frac{s_1}{r},\ldots,\frac{s_m}{r}\biggr) &\in \Cr_0, \label{eq:lex-proof-step1}
\intertext{and for any $\e > 0$,}
\frac{(1-\e) \tv_i}{s_i} &\in \Cr_i, \quad i \in [m].
\label{eq:lex-proof-step2}
\end{align}
\end{subequations}
Consequently,
\[
(1-\e) \biggl(\frac{\tv_1}{r},\ldots, \frac{\tv_m}{r}\biggr) \in
\Cr_0 \lexprod (\Cr_1,\ldots,\Cr_m).
\]
Since $\e > 0$ is arbitrary, this would establish the desired proof of the converse.

We now verify~\eqref{eq:lex-proof-steps} for an appropriate $(s_1,\ldots,s_m)$.
Let $Y = \phi(\Xv_1,\ldots,\Xv_m) \in \{0,1\}^r$ be the encoder output of the given index code
for independent and uniformly distributed messages,
which induces the joint distribution of the form
\begin{align}
\label{eq:induced-dist}
p(\xv_1, \ldots, \xv_m,y) = p(\xv_1) \cdots p(\xv_m)p(y|\xv_1,\ldots,\xv_m)
\end{align}
such that $Y$ is a function of $(\Xv_1,\ldots,\Xv_m)$ and
$X_j$ is a function of $(Y,X(A_j))$ for every $j \in V(G)$, namely, $X_j = \psi_j(Y,X(A_j))$, $j \in V(G)$.
Now let
\[
s_i = I(\Xv_i; Y | \Xv(A'_i)), \quad i \in [m],
\]
where the mutual information is evaluated under the joint distribution in~\eqref{eq:induced-dist}. Then, by Theorem~\ref{thm:capacity-it}
(with $U_i = \Xv_i$ and $V = Y$), we have~\eqref{eq:lex-proof-step1}.
For~\eqref{eq:lex-proof-step2},  we first state two lemmas, the proofs of which are
presented in Appendices~\ref{app:cap-property} and~\ref{app:mappings}.

\begin{lemma}
\label{lem:cap-property}
For any $n$-message index coding problem $(i|A_i)$, $i \in [n]$, with side information graph $G$, let
\[
(\phi(x^n),\psi_1(y,x(A_1)),\ldots,\psi_n(y,x(A_n)))
\]
be the encoder and decoders of a $(t_1,\ldots,t_n,r)$ index code under a \emph{relaxed} decoding condition
\[
\psi_i(\phi(x^n),x(A_i)) = x_i, \quad i \in J,
\]
for some subset $J \subseteq [n]$ of the messages. Then,
\[
\frac{(t_i: i\in J)}{r} \in \Cr(G|_J).
\]
\end{lemma}

\begin{lemma}
\label{lem:mappings}
Let $\e > 0$ and $s_i = I(\Xv_i; Y | \Xv(A'_i))$.
Then there exist mappings
\begin{subequations} \label{eq:mappings}
\begin{align}
\label{eq:mappings-1}
\phi'_i(\xv^k) &\in \{0,1\}^{k s_i/(1-\e)}, \quad i \in [m],
\intertext{and}
\psi'_j(w_i, x^k(A_j)) &\in \{0,1\}^{k t_j}, \quad j \in V(G_i),
\intertext{such that}
\psi'_j(\phi'_i(\xv^k), x^k(A_j)) &= x_j^k, \quad i \in [m],\, j \in V(G_i),
\end{align}
for $k$ sufficiently large.
\end{subequations}
\end{lemma}

Now we are ready to verify \eqref{eq:lex-proof-step2}.
We first apply Lemma~\ref{lem:mappings} for each $i \in [m]$.
The mappings $\phi'_i(\xv^k)$ and $\psi'_j(w_i, x^k(A_j))$, $j \in V(G_i)$,
form a $(k\tv_1,\ldots,k\tv_m,k s_i/(1-\e))$ index code for $G$ under the relaxed decoding condition that only $\xv_i^k = (x_j^k: j \in V(G_i))$ is required to be recovered correctly. Hence, by Lemma~\ref{lem:cap-property},
we can conclude that \eqref{eq:lex-proof-step2} holds.
This completes
the proof of Theorem~\ref{thm:cap-generallexprod}.

\section{Concluding Remarks}
\label{sec:conclusion}

The generalized lexicographic product structure investigated in this paper provides a natural method of building a larger index coding problem from smaller problems so that the capacity region of the larger problem can be expressed in the same generalized lexicographic product structure from the subproblem capacity regions. This leads to a divide-and-conquer approach to computing the index coding capacity region, either through direct decomposition (Theorem~\ref{thm:cap-generallexprod}) or by sandwiching between two generalized lexicographic products 
   (Corollary~\ref{coro:noncritical}).
   
Since the capacity region of a general $n$-message index coding problem is known for $n \le 5$, we can test this divide-and-conquer approach for all problems with six or less messages. Table~\ref{tb:glp-numbers} lists the number of all nonisomorphic $n$-message index coding problems $N$, along with the number of problems that are generalized lexicographic products of smaller graphs ($N_\mathrm{GLP}$), the number of problems that are sandwiched between two generalized lexicographic products of the same capacity region ($N_\mathrm{Sand}$), and the percentage of the problems whose capacity regions can be characterized by this \emph{divide-and-conquer} approach. This simple approach solves about one half of the 6-message problems without explicitly computing any inner and outer bounds on the capacity region.

\begin{table}[t]
\caption{The numbers of $n$-message index coding problems whose capacity regions can be characterized by the divide-and-conquer approach based on Theorem~\ref{thm:cap-generallexprod} and Corollary~\ref{coro:noncritical}.}
\label{tb:glp-numbers}
\centering
\begin{tabular}{c|c|c|c|c}
\hline
Number of messages & $N$ & $N_\mathrm{GLP}$ & $N_\mathrm{Sand}$ & $(N_\mathrm{GLP}+ N_\mathrm{Sand})/N$\\
\hline
2 & 3 & 3 & 0 & $100\%$ \\
3 & 16 & 11 & 3 & $87.5\%$ \\
4 & 218 & 110 & 70 & $82.6\%$ \\
5 & 9,608 & 2,511 & 4,054 & $68.3\%$ \\
6 & 1,540,944 & 161,989 & 607,161 & $49.9\%$ \\
\hline
\end{tabular}
\end{table}

Identifying the generalized lexicographic product structure in a general side information graph is
a computationally challenging problem (see Remarks~\ref{rmk:verify-gen-lex-prod} and \ref{rmk:verify-beyond-gen-lex-prod}). 
We offer the following
algorithmic questions that would shed some light on the current line of investigation:
\begin{itemize}
\item Given a graph $G$, can we efficiently determine whether $G$ is a generalized lexicographic product of smaller graphs? Although only a very small number of graphs are generalized lexicographic products, 
the capacity regions of many other graphs can be tightly sandwiched by the capacity regions of these graphs.

\item Can we efficiently transform a graph $G$ into a generalized lexicographic product by adding or removing a few edges? A recursive application of this procedure can yield a general outer or inner bound on the capacity region.
\end{itemize}
\section*{Acknowledgments}
The authors would like to thank the Associate Editor and anonymous reviewers
for their constructive comments, which improved the readability of the manuscript significantly.
They also would like to thank Elena Grigorescu and Minshen Zhu for pointing out an error in
an earlier proof of Theorem~1 based on the clique number of confusion graphs,
and acknowledge Parastoo Sadeghi and Jacques Verstraete for helpful discussions.
This work was supported by the National Science Foundation under Grant CCF-1320895
and the Korean Ministry of Science, ICT and Future Planning under the
Institute for Information and Communications Technology Promotion Grant B0132-15-1005
(Development of Wired-Wireless Converged 5G Core Technologies).

\appendices

\section{Proof of Lemma~\ref{lem:cap-property}}
\label{app:cap-property}

We construct an index code for problem $G|_J$ by setting $x_i = 0$, $i \not \in J$, in $\phi(x^n)$ and $\psi_i$, $i \in J$.
For every $x^n \in \Pi_{i = 1}^n \{0,1\}^{t_i}$,  define $\xt^n = \xt^n(x^n)$ by
\begin{align*}
\xt_i = \begin{cases}
x_i, & i \in J, \\
0, & i \not \in J,
\end{cases}
\end{align*}
represented in the same $t_i$ bits.
Note that the side information set of receiver $i \in J$ for problem $G|_J$ is $A_i \cap J$.
Let
\begin{align*}
\phi'(x(J)) = \phi(\xt^n) \in \{0,1\}^r,
\end{align*}
and
\begin{align*}
\psi'_i(y,x(A_i \cap J)) = \psi_i(y, \xt(A_i)).
\end{align*}
Then, by the given decoding condition,
\begin{align*}
\psi'_i(\phi'(x(J)),x(A_i \cap J)) = \psi_i(\phi(\xt^n),\xt(A_i)) = \xt_i = x_i, \quad i \in J.
\end{align*}
Hence, the mappings $\phi'(x(J))$ and $\psi_i'(y,x(A_i \cap J))$, $i \in J$, form a valid index code for the problem $G|_J$.
This completes the proof of the lemma.

\section{Proof of Lemma~\ref{lem:mappings}}
\label{app:mappings}

At a high level, the proof is based on random coding for rate--distortion theory~\cite{Shannon1959} and joint
typicality encoding \cite[Sec.~3.6]{El-Gamal--Kim2011} over $k$ copies of $(\Xv_1,\ldots,\Xv_m,Y)$.
For each $i \in [m]$, consider the joint distribution $p(\xv_i,\xv(A'_i),y)$ from \eqref{eq:induced-dist} and fix the conditional distribution $p(y|\xv(A'_i))$.
For each $\xv^k(A'_i)$, generate $kr$-bit sequences $y_i^k(w_i|\xv^k(A'_i))$, $w_i \in [2^{k s_i/(1-\e)}]$, each i.i.d.\@ according to $p(y|\xv(A'_i))$.
Then by the covering lemma \cite[Lemma 3.3]{El-Gamal--Kim2011}, with high probbability there exists at least one $w_i$ 
such that
\begin{align} \label{eq:joint-typicality}
(\xv_i^k, y_i^k(w_i|\xv^k(A'_i)), \xv^k(A'_i)) \in \aepk(\Xv_i,Y,\Xv(A'_i)),
\end{align}
provided that $k$ is sufficiently large and
\begin{align*}
s_i/(1-\e) > I(\Xv_i;Y | \Xv(A'_i)).
\end{align*}
If there is such a $w_i$ (if there is more than one, choose one arbitrarily), then we set 
\begin{align*}
\phi'_i(\xv^k) = w_i.
\end{align*}
Note by~\eqref{eq:joint-typicality} that the chosen $w_i$ is a function of $\xv_i^k$ and $\xv^k(A'_i)$ (and thus of $\xv^k$).
If there is no such index, set $\phi'_i(\xv^k) = 1$.

We now define $\psi'_j$ for each $j \in V(G_i)$. Let
\begin{align*}
\psi'_j(w_i,x^k(A_j)) = \psi_j(y_i^k(w_i|\xv^k(A'_i)), x^k(A_j)),
\end{align*}
where $\psi_j$ is the decoding function of the given index code for problem $G$ that is employed $k$ times.
Suppose that the joint typicality in~\eqref{eq:joint-typicality} holds among
$\xv_i^k$, $y_i^k(w_i|\xv^k(A'_i))$, and $\xv^k(A'_i)$. Then by the properties of joint typicality \cite[Section 2.5]{El-Gamal--Kim2011}, any functional relationship for them should hold, namely
\begin{align*}
x^k_j = \psi_j(y_i^k, x^k(A_j)) = \psi'_j(\phi'_i(\xv^k),x^k(A_j)), \quad j \in V(G_i).
\end{align*}
Therefore, as long as $w_i$ satisfying \eqref{eq:joint-typicality} is found, which happens with high probability,
the mappings $\phi'_i$ and $\psi'_j$ defined above satisfy the desired properties in \eqref{eq:mappings} with high probability.
Finally, by Remark~\ref{rmk:epsilon-error}, we can come up with mappings for which these properties hold 
for every sequence with a negligible decrease in the rates. This completes the proof of the lemma.

\bibliographystyle{IEEEtran}
\bibliography{nit}

\end{document}

%% file: defns.tex
\usepackage{xspace}
\usepackage{bbm}
\input{mathlig}

\newcommand{\muspace}{\mspace{1mu}}

\DeclareRobustCommand{\scond}{\mathchoice{\muspace\vert\muspace}{\vert}{\vert}{\vert}}
\mathlig{|}{\scond}

\DeclareRobustCommand{\discint}{\mathchoice{\mspace{-1.5mu}:\mspace{-1.5mu}}{\mspace{-1.5mu}:\mspace{-1.5mu}}{:}{:}}
\mathlig{::}{\discint}
\newcommand{\suchthat}{\mathchoice{\colon}{\colon}{:\mspace{1mu}}{:}}

\newcommand{\Ec}{\mathcal{E}}

\newcommand{\Uc}{\mspace{1.5mu}\mathcal{U}}
\newcommand{\Vc}{\mathcal{V}}

\newcommand{\Cr}{\mathscr{C}}

\newcommand{\Xv}{{\bf X}}

\newcommand{\Rv}{{\bf R}}

\newcommand{\av}{{\bf a}}
\newcommand{\bv}{{\bf b}}

\newcommand{\tv}{{\bf t}}
\newcommand{\xv}{{\bf x}}

\newcommand{\betav}{{\boldsymbol \beta}}

\newcommand{\rhov}{{\boldsymbol \rho}}

\newcommand{\aepk}{{\mathcal{T}_{\epsilon}^{(k)}}}

\newcommand{\Xh}{{\hat{X}}}

\newcommand{\xh}{{\hat{x}}}

\newcommand{\xt}{{\tilde{x}}}

\def\b{\beta}

\def\d{\delta}
\def\e{\epsilon}

\let\P\relax
\DeclareMathOperator\P{\textsf{P}}

\def\textiid{i.i.d.\@\xspace}
\newcommand\iid{\ifmmode\text{ i.i.d. } \else \textiid \fi}

\def\mathllap{\mathpalette\mathllapinternal}
\def\mathllapinternal#1#2{%
  \llap{$\mathsurround=0pt#1{#2}$}}

\def\clap#1{\hbox to 0pt{\hss#1\hss}}
\def\mathclap{\mathpalette\mathclapinternal}
\def\mathclapinternal#1#2{%
  \clap{$\mathsurround=0pt#1{#2}$}}

\let\oldstackrel\stackrel
\renewcommand{\stackrel}[2]{\oldstackrel{\mathclap{#1}}{#2}}

\renewcommand{\hbar}{h\mathllap{\overline{\vphantom{h}\hphantom{\rule{4.6pt}{0pt}}}\mspace{0.77mu}}}

\catcode`~=11 
\newcommand{\urltilde}{\kern -.06em\lower -.06em\hbox{~}\kern .02em}
\catcode`~=13 

%% file: mathlig.tex
\catcode`@=11
\chardef\mathlig@atcode\count255

\def\actively#1#2{\begingroup\uccode`\~=`#2\relax\uppercase{\endgroup#1~}}
\def\mathlig@gobble{\afterassignment\mathlig@next@cmd\let\mathlig@next= }
\def\mathlig@delim{\mathlig@delim}
\def\mathlig@defcs#1{\expandafter\def\csname#1\endcsname}
\def\mathlig@let@cs#1#2{\expandafter\let\expandafter#1\csname#2\endcsname}
\def\mathlig@appendcs#1#2{\expandafter\edef\csname#1\endcsname{\csname#1\endcsname#2}}

\def\mathlig#1#2{\mathlig@checklig#1\mathlig@end\mathlig@defcs{mathlig@back@#1}{#2}\ignorespaces}

\def\mathlig@checklig#1#2\mathlig@end{%
 \expandafter\ifx\csname mathlig@forw@#1\endcsname\relax
 \expandafter\mathchardef\csname mathlig@back@#1\endcsname=\mathcode`#1%
 \mathcode`#1"8000\actively\def#1{\csname mathlig@look@#1\endcsname}%
 \mathlig@dolig#1\mathlig@delim
\fi
\mathlig@checksuffix#1#2\mathlig@end
}

\def\mathlig@checksuffix#1#2\mathlig@end{%
\ifx\mathlig@delim#2\mathlig@delim\relax\else\mathlig@checksuffix@{#1}#2\mathlig@end\fi
}
\def\mathlig@checksuffix@#1#2#3\mathlig@end{%
\expandafter\ifx\csname mathlig@forw@#1#2\endcsname\relax\mathlig@dosuffix{#1}{#2}\fi
\mathlig@checksuffix{#1#2}#3\mathlig@end
}

\def\mathlig@dosuffix#1#2{%
\mathlig@appendcs{mathlig@toks@#1}{#2}%
\mathlig@dolig{#1}{#2}\mathlig@delim
}

\def\mathlig@dolig#1#2\mathlig@delim{%
 \mathlig@defcs{mathlig@look@#1#2}{%
 \mathlig@let@cs\mathlig@next{mathlig@forw@#1#2}\futurelet\mathlig@next@tok\mathlig@next}%
 \mathlig@defcs{mathlig@forw@#1#2}{%
  \mathlig@let@cs\mathlig@next{mathlig@back@#1#2}%
  \mathlig@let@cs\checker{mathlig@chck@#1#2}%
  \mathlig@let@cs\mathligtoks{mathlig@toks@#1#2}%
  \expandafter\ifx\expandafter\mathlig@delim\mathligtoks\mathlig@delim\relax\else
  \expandafter\checker\mathligtoks\mathlig@delim\fi
  \mathlig@next
 }%
 \mathlig@defcs{mathlig@toks@#1#2}{}%
 \mathlig@defcs{mathlig@chck@#1#2}##1##2\mathlig@delim{%
  \ifx\mathlig@next@tok##1%
   \mathlig@let@cs\mathlig@next@cmd{mathlig@look@#1#2##1}\let\mathlig@next\mathlig@gobble
  \fi 
  \ifx\mathlig@delim##2\mathlig@delim\relax\else
   \csname mathlig@chck@#1#2\endcsname##2\mathlig@delim
  \fi
 }%
 \ifx\mathlig@delim#2\mathlig@delim\else
  \mathlig@defcs{mathlig@back@#1#2}{\csname mathlig@back@#1\endcsname #2}%
 \fi
}%

\catcode`@\mathlig@atcode